\documentclass[letterpaper]{IEEEtran}

\ifCLASSINFOpdf
\else
   \usepackage[dvips]{graphicx}
\fi
\usepackage{url}

\usepackage{graphicx}
\usepackage{pdfpages}
\usepackage{setspace}
\usepackage{multirow}
\usepackage{amsmath}
\usepackage{epsfig}
\usepackage{array}
\usepackage{color}
\usepackage{subfigure}
\usepackage{xcolor}
\usepackage{amssymb}
\usepackage{url}
\usepackage{mathrsfs}
\usepackage{booktabs}
\usepackage{cite}
\usepackage{bbding}
\usepackage{upgreek}
\usepackage{amssymb}
\usepackage{amsfonts}
\usepackage[noend]{algpseudocode}
\usepackage{algorithmicx,algorithm}
\renewcommand{\baselinestretch}{0.958}

\begin{document}
\title{Glance and Gaze: A Collaborative Learning Framework for Single-channel Speech Enhancement}
\author{Andong~Li,
	Chengshi~Zheng,
	Lu~Zhang,
	and Xiaodong~Li
	\thanks{Andong~Li, Chengshi~Zheng, and Xiaodong Li are with the Key Laboratory of Noise and Vibration Research, Institute of Acoustics, Chinese Academy of Science,
		Beijing, 100190, China, and also with University of Chinese Academy of Sciences, Beijing, 100049, China. (email:\{liandong, cszheng, lxd\}cszheng@mail.ioa.ac.cn).}% <-this % stops a space
	\thanks{Lu Zhang is with the Department of Electronics and Information Engineering, Harbin Institute of Technology, Shenzhen, Shenzhen 518000, China. (email:18B952047@stu.hit.edu.cn).}
	\thanks{This work was supported by NSFC
		(National Science Fund of China) under Grant No. 61571435.}
	\thanks{Manuscript received July XX, 2017; revised XXXX XX, XX.}}

\markboth{Journal of \LaTeX\ Class Files, Vol. 14, No. 8, August 2015}
{Shell \MakeLowercase{\textit{et al.}}: Bare Demo of IEEEtran.cls for IEEE Journals}
\maketitle
\begin{abstract}
The capability of the human to pay attention to both coarse and fine-grained regions has been applied to computer vision tasks. Motivated by that, we propose a collaborative learning framework in the complex domain for monaural noise suppression. The proposed system consists of two principal modules, namely spectral feature extraction module (FEM) and stacked glance-gaze modules (GGMs). In FEM, the UNet-block is introduced after each convolution layer, enabling the feature recalibration from multiple scales. In each GGM, we decompose the multi-target optimization in the complex spectrum into two sub-tasks. Specifically, the glance path aims to suppress the noise in the magnitude domain to obtain a coarse estimation, and meanwhile, the gaze path attempts to compensate for the lost spectral detail in the complex domain. The two paths work collaboratively and facilitate spectral estimation from complementary perspectives. Besides, by repeatedly unfolding the GGMs, the intermediate result can be iteratively refined across stages and lead to the ultimate estimation of the spectrum. The experiments are conducted on the WSJ0-SI84, DNS-Challenge dataset, and Voicebank+Demand dataset. Results show that the proposed approach achieves state-of-the-art performance over previous advanced systems on the WSJ0-SI84 and DNS-Challenge dataset, and meanwhile, competitive performance is achieved on the Voicebank+Demand corpus.
\end{abstract}

\begin{IEEEkeywords}
%Enter key words or phrases in alphabetical order, separated by commas. For a list of suggested keywords, send a blank e-mail to keywords@ieee.org or visit \url{http://www.ieee.org/organizations/pubs/ani_prod/keywrd98.txt}
Monaural speech enhancement, glance and focus, complex domain, multi-task learning.
\end{IEEEkeywords}

\IEEEpeerreviewmaketitle
\vspace{-0.35cm}
\section{Introduction}
%\vspace{-0.1cm}
\IEEEPARstart{E}NVIRONMENTAL noise is one of the major factors to hinder the application of modern hands-free communication systems and hearing assistant devices. Monaural speech enhancement (SE) aims to extract the target speech from the noisy mixture and improve the speech quality and intelligibility when only a single-channel recording is available~{\cite{loizou2013speech}}. Recent years have witnessed the renaissance of deep neural networks (DNNs), and a plethora of DNN-based SE algorithms have been proposed, which are demonstrated to be successful under highly non-stationary noise environments and extremely low signal-to-noise ratio (SNR) conditions~{\cite{wang2018supervised, xu2013experimental, siniscalchi2021vector}}.
	
In the time-frequency (T-F) domain, the estimation of magnitude is often investigated and incorporated with the noisy phase~{\cite{ephraim1984speech, tan2018gated}}. This is because phase information usually exhibits nonstructural regularity and it is still a challenging task to manipulate the phase spectrum accurately~{\cite{gerkmann2015phase, mowlaee2016advances}}. Recently, the importance of phase in improving perception quality under low SNRs has been emphasized~{\cite{paliwal2011importance}} and a multitude of phase-aware DNN-based SE algorithms are proposed henceforth, which can be roughly divided into two categories, namely, complex-domain~{\cite{williamson2015complex, tan2020learning, lee2019joint}} and time-domain~{\cite{pascual2017segan, luo2019conv, pandey2019new, defossez2020real}}. For the former, the magnitude and phase are coupled into real and imaginary (RI) components in the complex domain and the phase can be implicitly estimated by mapping to the clean speech RI components. For the latter, the waveform serves as both the input and output, which diverts around the explicit phase estimation. However, since scale-invariant SNR (SI-SNR)~{\cite{le2019sdr}} or mean square error (MSE) is usually utilized as the loss function in the time domain, the subjective speech quality cannot be guaranteed as the criterion correlates poorly with the human ratings~{\cite{reddy2020dnsmos}}. Hence, this study still focuses on SE research in the complex domain.
	
More recently, the multi-stage concept begins to thrive in the SE area~{\cite{gao2016snr, li2021icassp, hao2020masking}}. Because it has been shown that the intermediate priors can boost the subsequent optimization by decomposing the original task into multiple sub-tasks. This is different from the previous single-stage paradigm, in which the mapping process is often packed into only one black box, resulting in poor interpretability assessments. In~{\cite{gao2016snr}}, multiple SNR-improved sub-targets were set and trained with a light-weight module. However, as each sub-network is coerced to improve the SNR by a small margin, the performance upper-bound is strictly restricted. In~{\cite{li2021icassp, hao2020masking}}, a two-stage pipeline was designed, where a coarse spectrum is first obtained, followed by the refinement module in the second stage to further polish the spectral detail. However, the shortage of such a pipeline attributes to its sequential nature, \emph{i.e.}, the performance of the second stage heavily depends on the previous output. In other words, such cascaded dependencies require that the second-stage model should have sufficient tolerance to rectify the estimation error caused by the previous stages.   
 %In other words, the erroneous pre-estimation may not be rectified by the later refinement module.

\begin{figure*}[t]
	\centering
	\centerline{\includegraphics[width=1.62\columnwidth]{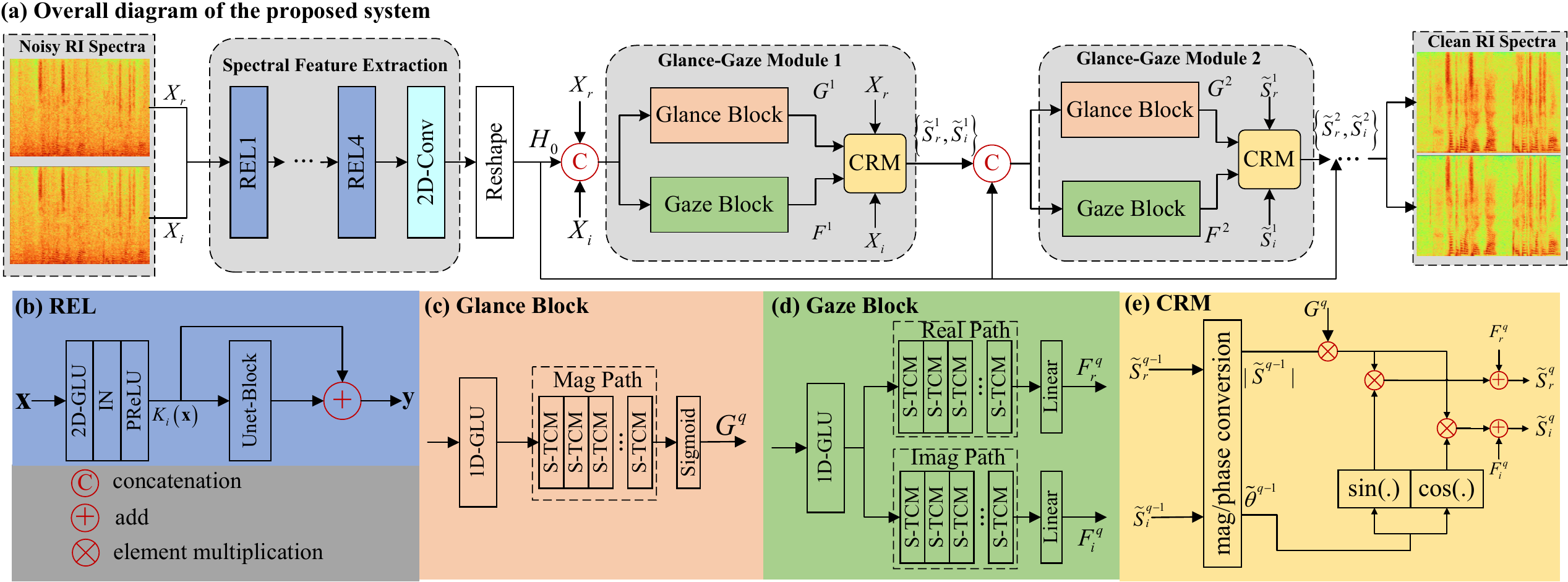}}
	\caption{Architecture of the proposed GaGNet. (a) The overall diagram of the proposed system. (b) The detail of REL. (c) The detail of the Glance Block. (d) The detail of the Gaze Block. (e) The detail of the CRM.}
	\label{fig:diagram-system}
	\vspace{-0.4cm}
\end{figure*}

%To imitate the behavior of mankind in knowledge acquisition from both global and local scales simultaneously, we present a novel framework named \textbf{G}lance \textbf{a}nd \textbf{F}ocus Network (GaFNet), where the speech components are extracted from complementary perspectives in a collaborative method.

In this letter, we propose a parallel structure for coarse and refined estimation respectively, enabling hierarchical optimization toward the complex spectrum. This newly proposed framework imitates the behaviour of mankind in knowledge acquisition from global and local scales simultaneously~{\cite{wang2020glance}}, and we name it \textbf{G}lance \textbf{a}nd \textbf{G}aze Network (GaGNet). Specifically, we propose a dual-path structure, where the glance path aims to recover the magnitude component by noise filtering, and the gaze path is tasked with the complex spectral detail inpainting. The rationale is two-fold. First, Li \emph{et.al.}~{\cite{li2021icassp}} revealed the significance to decouple magnitude and phase optimization, so we separately consider the magnitude and phase optimization from two parallel paths. Besides, there often exist residual noise components and meanwhile, some spectral details tend to get lost. Therefore, the introduction of the gaze path can further suppress the remaining distortion and meanwhile, repair the missing details. In this study, we also inherit the merits of multi-stage training by repeatedly stacking multiple modules to facilitate the spectrum recovery in a stage-wise manner.
\vspace{-0.4cm}
\section{Methodlogy}
\label{methodology}
\vspace{-0.1cm}
The overall diagram of the proposed system is shown in Fig.~{\ref{fig:diagram-system}}(a). It is mainly comprised of two parts, namely the spectral feature extraction module (FEM) and multiple Glance-Gaze modules (GGMs). As multiple GGMs are stacked, The number of GGMs is denoted as \textit{Q}. The network input is the noisy complex spectrum, denoted as $X = Cat\left(X_{r}, X_{i}\right) \in \mathbb{R}^{2\times T \times F}$, and the clean target is denoted as $S = Cat\left(S_{r}, S_{i}\right)\in \mathbb{R}^{2\times T \times F}$, where $Cat$ denotes the concatenation operation, and $T$, $F$ denote the size in the time and frequency axes, respectively. Subscripts $r$ and $i$ namely denote the real and imaginary parts. We will illustrate each module in detail below.
\vspace{-0.4cm}
\subsection{Spectral Feature Extraction}
\label{spectral-feature-extraction}
\vspace{-0.1cm}
In the previous studies~{\cite{tan2020learning, li2021icassp}}, the UNet-style encoder was generally adopted as the feature extractor, where multiple plain two-dimension convolution layers (2D-Convs) or their gated liear unit (GLU) formats~{\cite{dauphin2017language}} are stacked to gradually compress the feature map. However, since the small kernel size is utilized, only local contextual information can be learned. In addition, consecutive downsampling operations will cause spectral information loss. Inspired by U$^{2}$Net~{\cite{qin2020u2}}, we replace the 2D-Convs by our proposed recalibrate encoder layers (RELs) to mitigate the information loss while enlarging the feature scale, which are shown in Fig.~{\ref{fig:diagram-system}}(b). It mainly consists of a 2D-GLU, instance normalization (IN), PReLU~{\cite{he2015delving}}, and a UNet-Block with the residual connection. For UNet-Block, it receives the current feature map $\mathcal{K}_{i}(\mathbf{x})$ after 2D-GLU as the input and then further encodes the contextual information with a UNet-style structure~{\cite{qin2020u2}}. The rationale is two-fold. First, the UNet-Block can grasp multi-scale information, which enhances the feature representation capability. Besides, the encoder layers close to the input are usually noisy and the UNet-Block can further recalibrate the feature distribution. The process can be given by:
\vspace{-0.1cm}
\begin{gather}
\label{eqn1}
\mathcal{K}_{i}\left(\mathbf{x}\right) = GLU\left( \mathbf{x}  \right),\\
\mathbf{\mathbf{y}} = UNet\text{-}Block\left( \mathcal{K}_{i}(\mathbf{x}) \right) + \mathcal{K}_{i}(\mathbf{x}),
\end{gather}
where $i\in\left\{ 1,\cdots 4 \right\}$ is the layer index, \emph{i.e.}, 4 RELs are employed herein.
\vspace{-0.5cm}
\subsection{Glance-Focus Module}
\label{glance-focus-module}
\vspace{-0.1cm}
After FEM, the obtained feature map is denoted as $H_{0}\in \mathcal{R}^{C\times T \times F'}$. Note that we keep the time dimension to preserve the temporal resolution. Then the feature map is reshaped into ${C'\times T}$, where $C' = C \times F'$, followed by multiple GGMs. The motivation of GGM is induced by the physiological phenomenon that human can pay attention to both global and local components without much effort. Therefore, two parallel blocks are designed accordingly, namely Glance Block (GLB) and Gaze Block (GAB). In the GLB, the output is a gain function to suppress the noise in the magnitude domain, leading to the coarse ``glance'' toward the overall spectrum. For the GAB, it seeks to predict the residual to repair the spectral detail in the complex domain, which serves as the ``gaze'' operation. Both outputs are applied to the collaborative reconstruction module (CRM) to obtain the spectrum estimation. In addition, we adopt the multi-stage training strategy, \emph{i.e.}, multiple GGMs are repeatedly stacked and the RI components of current stage are iteratively updated based on that of the last stage.

The inputs of GGM are from two sources, namely $H_{0}$ and the estimated RI from the last stage, \emph{i.e.}, $Cat\left( H_{0}, \tilde{S}^{q-1}_{r}, \tilde{S}^{q-1}_{i} \right) \in \mathbb{R}^{(C'+2F)\times T}$. Note that in the first GGM, the input RI denotes the original version. GLB and GAB have the similar topology structure, as shown in Fig.~{\ref{fig:diagram-system}}(c)-(d), where the 1D-GLU first compresses the input features and then multiple temporal convolutional modules (TCMs)~{\cite{luo2019conv}} are leveraged to model the long-range temporal dependencies. As GLB is operated to obtain the mask in the magnitude domain, we only use single-path TCMs, and the sigmoid function is utilized to scale the value into $(0, 1)$. For GAB, dual-path TCMs are adopted to capture RI spectral details and linear layers are used as the output layer. Note that we adopt a squeezed version of TCM herein, \emph{i.e.}, S-TCM, which dramatically alleviates the parameter burden~{\cite{zhang2020deepmmse}}.

Let $G^{q}$ and $F^{q} = \left\{  F^{q}_{r}, F^{q}_{i}    \right\}$ be the output of GLB and GAB, respectively, they are sent to CRM together with the previous estimation to update the RI components of current stage in a collaborative manner, which is the core step of the whole framework, as shown in Fig.~{\ref{fig:diagram-system}}(e). To be specific, in the $q${th} GGM, the input RI spectra $\left\{ \tilde{S}^{q-1}_{r}, \tilde{S}^{q-1}_{i}  \right\}$ are first decoupled into magnitude and phase, given by:
\vspace{-0.2cm}
\begin{equation}
\label{eqn3}
\lvert \tilde{S}^{q-1} \rvert = \sqrt{ \lvert  \tilde{S}^{q-1}_{r} \rvert^2 + \lvert  \tilde{S}^{q-1}_{i} \rvert^2  },\\
\end{equation}
\begin{equation}
\label{eqn4}
\angle \tilde{\theta}^{q-1} = \arctan \left( \tilde{S}^{q-1}_{r}, \tilde{S}^{q-1}_{i}  \right),
\end{equation}
\vspace{-0.4cm}

The magnitude part is then multiplied with $G^q$, whose motivation is to suppress the noise coarsely. However, the spectral detail also needs to be considered. As such, $F^{q}$ is designed to focus on the lost detail from the complex-domain perspective. In a nutshell, the whole procedure is thus formulated as:
%\vspace{-0.2cm}
\begin{equation}
\label{eqn5}
\lvert\tilde{S}^{q, fil}\rvert = \lvert\tilde{S}^{q}\rvert \otimes G^{q},
\end{equation}
\begin{equation}
\label{eqn6}
\tilde{S}^{q, fil}_{r} = \vert\tilde{S}^{q, fil}\rvert \otimes \cos\left( \tilde{\theta}^{q-1} \right),
\end{equation}
\begin{equation}
\label{eqn7}
\tilde{S}^{q, fil}_{i} = \lvert\tilde{S}^{q, fil}\rvert \otimes \sin\left( \tilde{\theta}^{q-1} \right),
\end{equation}
\begin{equation}
\label{eqn8}
\tilde{S}^{q}_{r} = \tilde{S}^{q, fil}_{r} + F^{q}_{r},
\end{equation}
\begin{equation}
\label{eqn9}
\tilde{S}^{q}_{i} = \tilde{S}^{q, fil}_{i} + F^{q}_{i},
\end{equation}
where superscript $(\cdot)^{fil}$ denotes the filtered magnitude. $\otimes$ refers to the element-wise multiplication operator.    
\vspace{-0.5cm}
\subsection{Loss Function}
\label{loss-function}
\vspace{-0.1cm}
As stated above, GGM will generate the estimated RI spectra in each intermediate stage. Similar to~{\cite{gao2016snr}}, we adopt a weighted multi-target MSE loss, given by:
\vspace{-0.2cm}
\begin{equation}
\label{eqn10}
\mathcal{L} = \sum_{q=1}^{Q} \lambda_{q}\mathcal{L}_{q},
\end{equation}
where $\lambda_{q}$ is the weighted coefficient in the $q^{th}$ stage. In this context, $\lambda_{Q} = 1.0$ and $\lambda_{q} = 0.1$ for $q \ne Q$ are chosen. $\mathcal{L}_{q}$ is given by:
\vspace{-0.0cm}
\begin{equation}
\label{eqn11}
\mathcal{L}_{q} = \frac{1}{2}\left(\left \| \tilde{S}^{q}_{r} - S_{r} \right \|^{2}_{F} + \left \| \tilde{S}^{q}_{i} - S_{r} \right \|^{2}_{F} + \left \| \lvert \tilde{S}^{q}\rvert - \lvert S \rvert \right\|^{2}_{F}\right),
\end{equation}

From Eq.{\ref{eqn11}}, one can find that except for RI loss, the magnitude constraint is also introduced as it is reported to improve the speech quality~{\cite{wang2020complex, wisdom2019differentiable}}. In addition, instead of providing the supervision toward the coarse and residual estimation explicitly, the supervision is only available toward their sum in each module, \emph{i.e.}, the estimated gain and residual are adaptively learned in the training process, resulting in the end-to-end training.
\vspace{-0.2cm}
\section{Experimental Setup}
\label{experimental-setup}
\vspace{-0.1cm}
\subsection{Dataset}
\label{dataset}
\vspace{-0.1cm}
To evaluate the performance of the proposed system, we conduct the experiments on the WSJ0-SI84 corpus~{\cite{paul1992design}}, which consists of 7138 clean utterances by 83 speakers (42 males and 41 females). We randomly choose 5,428 training utterances and 957 validation utterances from 77 speakers, respectively. In addition, 150 utterances from 6 untrained speakers are selected for testing. For the noise set, around 20,000 noises are selected from DNS-Challenge noise set~{\cite{reddy2020interspeech}} for training, whose duration is around 55 hours. In addition, two challenging untrained noises are employed to demonstrate the model generalization capability, namely babble, and factory1 from NOISEX92~{\cite{varga1993assessment}}.

During the mixing process, a random noise cut is extracted and then mixed with a randomly sampled utterance under a SNR selected from -5\rm{dB} to 0\rm{dB}. As a result, we totally generate around 150,000 and 10,000 noisy-clean pairs for training and validation (around 300 hours). For model evaluation, four SNR cases are set, \emph{i.e.}, $\left\{-3\rm{dB}, 0\rm{dB}, 3\rm{dB}, 6\rm{dB}\right\}$, and 300 noisy-clean pairs are generated for each case.
\vspace{-0.4cm}
\subsection{Baseline}
\label{baseline}
\vspace{-0.1cm}
Five advanced approaches are utilized as the baseline systems, namely GCRN~{\cite{tan2020learning}}, DCCRN~{\cite{hu2020dccrn}}, PHASEN~{\cite{yin2020phasen}}, AECNN~{\cite{pandey2019new}}, and ConvTasNet~{\cite{luo2019conv}}. GCRN, DCCRN, and PHASEN belong to the complex-domain family, and both magnitude and phase recovery are investigated. AECNN and ConvTasNet are two advanced time-domain models, where both input and target are raw time-domain waveforms. We reimplement the baselines following the best configuration mentioned in the literature except several modifications are given. First, for both GCRN and DCCRN, we adopt the RI loss+magnitude constraint to improve the speech quality. For AECNN, the frame size is 16384 samples with 50\% overlap, \emph{i.e.}, around 1-second context can be exploited for each frame. For ConvTasNet, it is originally used under 8kHz sampling rate, and we extend it to 16kHz for a fair comparison. Note that for PHASEN and AECNN, both past and future information are utilized while other systems are causal.
\vspace{-0.2cm}
\subsection{Implementation Details}
\label{implementaion-detail}
\vspace{-0.0cm}
\subsubsection{Model Details}
In the REL, the kernel size and stride of 2D-GLUs are $(2, 3)$ and $(1, 2)$ in the time and frequency axes, and the number of channels $C$ remains 64. For UNet-Block, the resultant values \emph{w.r.t.} kernel size and stride are $(1, 3)$ and $(1, 2)$. Set the (de)encoder layers in the UNet-Block as $M$, then $M_{i}$ is $\left\{4, 3, 2, 1\right\}$ for the four RELs. In the GGM, the compressed feature size \textit{D} is set to 256 after 1D-GLU. For each temporal modeling path, \textit{P} groups of S-TCMs are deployed, where each group includes 4 TCMs with kernel size and dilation rate being 3 and $\left\{1, 2, 5, 9\right\}$, respectively.
\vspace{-0.0cm}
\subsubsection{Training Details}
All the utterances are sampled at 16kHz and chunked to 8 seconds. The 20ms Hanning window is utilized with 50\% overlap between adjacent frames. 320-point FFT is used, leading to 161-D features. More recently, the efficacy of the power-compression strategy has been illustrated in the dereverberation task~{\cite{li2021importance}}, so we conduct the power compression toward the magnitude while leaving the phase unaltered, and the compression variable $\beta = 0.5$, which is reported a promising choice, \emph{i.e.}, $Cat\left(\lvert X \rvert^{0.5}\cos\left({\theta_{X}}\right), \lvert X \rvert^{0.5}\sin\left({\theta_{X}}\right)\right)$, $Cat\left(\lvert S \rvert^{0.5}\cos\left({\theta_{S}}\right), \lvert S \rvert^{0.5}\sin\left({\theta_{S}}\right)\right)$. All models are optimized using Adam~{\cite{kingma2014adam}} with the learning rate of 0.0005, the batch size of 8 at the utterance level, and the epoch number of 60.

\renewcommand\arraystretch{0.85}
\begin{table}[t]
	\caption{Results comparison with different \textit{P} and \textit{Q} configurations. All the values are averaged among different testing SNRs. \textbf{BOLD} denotes the best result in each case.}
	\footnotesize
	\setlength{\tabcolsep}{3pt}
	\centering
	%\resizebox{0.95\columnwidth}{!}{
		\begin{tabular}{ccccccc}
			\toprule
			\multirow{2}*{Config.} &\multicolumn{6}{c}{\textit{(P, Q)}}\\
			\cmidrule(lr){2-7}
			     & (1,3) &(2, 3) &(3, 3) &(2, 1) &(2, 2) &(2, 4)\\
			 \midrule
			 \#Param.(M) &4.31 &5.94 &7.58 &\textbf{2.33} &4.14 &7.76 \\
			 PESQ  &2.76 &2.84 &2.85 &2.76 &2.80 &\textbf{2.86} \\
			 ESTOI(\%) &75.62 &76.57 &76.96 &74.38 &75.93 &\textbf{77.05}\\
			 SDR(\rm{dB}) &11.33 &11.59 &11.71 &11.01 &11.44 &\textbf{11.73} \\
			\bottomrule
	\end{tabular}
	\label{tbl:ablation1}
	\vspace{-0.4cm}
\end{table}

%\renewcommand\arraystretch{0.95}
%\begin{table}[t]
%	\caption{Results comparison with different \textit{P} and \textit{Q} configurations. All the values are averaged among different testing SNRs. \textbf{BOLD} denotes the best result in each case.}
%	\tiny
%	\centering
%	\resizebox{0.89\columnwidth}{!}{
%		\begin{tabular}{c|c|ccc}
%			\hline
%			\textit{P} &\textit{Q} &PESQ &ESTOI(\%) &SDR(\rm{dB})\\
%			\hline
%			1 &3 &2.76 &75.62 &11.33\\
%			2 &3 &2.84 &76.57  &11.59  \\
%			3 &3 &\textbf{2.85}  &\textbf{76.96}  &\textbf{11.71}  \\
%			\hline
%			2 &1 &2.76  &74.38 &11.01 \\
%			2 &2  &2.80  &75.93  &11.44 \\
%			2 &4 &2.86  &77.05  &11.73 \\
%			2 &5 &\textbf{2.89}  &\textbf{77.36} &\textbf{11.78}  \\
%			\hline
%	\end{tabular}}
%	\label{tbl:ablation1}
%	\vspace{-0.5cm}
%\end{table}

\renewcommand\arraystretch{0.90}
\begin{table}[t]
	\caption{Results comparison with different spectral reconstruction algorithms.}
	\footnotesize
	\setlength{\tabcolsep}{3pt}
	\centering
	%\resizebox{0.95\columnwidth}{!}{
		\begin{tabular}{c|cccc}
			\hline
			Algorithms  &\#Param.(M) &PESQ &ESTOI(\%) &SDR(\rm{dB})\\
			\hline
			Mag-RM  &\textbf{2.63} &2.66 &70.17  &9.57\\
			Com-RM  &4.73 &2.76  &74.41 &11.18 \\
			Phasen-RM  &5.94 &2.76  &73.83 &10.74 \\
			CRM(Pro.) &5.94 &\textbf{2.84}  &\textbf{76.57}  &\textbf{11.59} \\
			\hline
	\end{tabular}
	\label{tbl:ablation2}
	\vspace{-0.4cm}
\end{table}

\renewcommand\arraystretch{0.70}
\begin{table*}[t]
	\caption{Objective result comparisons among different models in terms of PESQ, ESTOI and SDR in the test set. \textbf{Cau.} denotes whether to use the causal setup.}
	\small
	\setlength{\tabcolsep}{3pt}
	\centering
	\resizebox{0.98\textwidth}{!}{
		\begin{tabular}{ccccccccccc|ccccc|ccccc}
			\hline
			Metrics &\multirow{2}*{\rotatebox{90}{\textbf{Cau.}}}
			&\multirow{1}*{\#Param.} &\multirow{1}*{MACs}  &\multirow{1}*{Process} &\multirow{1}*{Memory} &\multicolumn{5}{c}{PESQ} &\multicolumn{5}{c}{ESTOI(\%)} &\multicolumn{5}{c}{SDR(dB)}\\
			%\cline{5-19}
			\cmidrule(lr){1-1}\cmidrule(lr){7-11}\cmidrule(lr){12-16}\cmidrule(lr){17-21}
			SNR(dB) & &(M) &(G/s) &time(s) &footprint(GB) &-3 &0 &3 &6 &\multicolumn{1}{c}{Avg.}  &-3 &0 &3  &6 &\multicolumn{1}{c}{Avg.} &-3 &0 &3 &6 &\multicolumn{1}{c}{Avg.}\\
			\cline{1-21}
			\multicolumn{1}{c}{Noisy} &- &- &- &- &- &1.60 &1.77 &1.99 &2.19 &\multicolumn{1}{c}{1.89}  &31.59 &40.23 &49.60 &58.64 &\multicolumn{1}{c}{45.02}  &-2.94 &0.04 &3.04 &6.03 &\multicolumn{1}{c}{1.54}\\
			
			\multicolumn{1}{c}{GCRN} &\Checkmark &9.77 &2.47 &0.20 &0.22 &2.32 &2.62 &2.87 &3.08 &\multicolumn{1}{c}{2.72}  &59.57 &68.83 &75.72 &80.78 &\multicolumn{1}{c}{71.23}  &6.39 &8.69 &10.71 &12.37  &\multicolumn{1}{c}{9.54}\\
			
			\multicolumn{1}{c}{DCCRN} &\Checkmark &\textbf{3.67} &11.13 &0.44 &0.67 &2.31 &2.61 &2.88 &3.10 &\multicolumn{1}{c}{2.72}  &59.57 &68.11 &75.86 &81.56 &\multicolumn{1}{c}{71.00}  &6.52 &9.08 &11.53 &13.60  &\multicolumn{1}{c}{10.18}\\
			
			\multicolumn{1}{c}{PHASEN} &\XSolidBrush &8.76 &6.12 &0.37 &0.93 &2.36 &2.70 &2.99 &3.21 &\multicolumn{1}{c}{2.82}  &61.78 &71.25 &78.32 &83.31 &\multicolumn{1}{c}{73.67}  &7.47 &9.83 &12.03 &13.82 &\multicolumn{1}{c}{10.79}\\
			
			\multicolumn{1}{c}{AECNN} &\XSolidBrush &6.65 &1.74 &\textbf{0.08} &\textbf{0.13} &2.32 &2.64 &2.90 &3.11 &\multicolumn{1}{c}{2.74}  &62.13 &71.57 &78.25 &83.03 &\multicolumn{1}{c}{73.74}  &7.74 &10.14 &12.29 &14.12  &\multicolumn{1}{c}{11.07}\\
			
			\multicolumn{1}{c}{ConvTasNet} &\Checkmark  &5.00 &5.23 &0.71 &2.13 &2.26 &2.52 &2.76 &2.96 &\multicolumn{1}{c}{2.63}  &63.88 &72.21 &78.49 &83.22 &\multicolumn{1}{c}{74.45}  &7.77 &10.05 &12.19 &14.06  &\multicolumn{1}{c}{11.02}\\
			\hline
			\multicolumn{1}{c}{GaGNet$^{\dagger}$(Pro.)}  &\Checkmark &5.62 &\textbf{0.92} &0.11 &0.23 &2.41 &2.71 &2.97 &3.19 &\multicolumn{1}{c}{2.82}  &65.55 &73.87 &80.01 &84.50 &\multicolumn{1}{c}{75.99}  &8.27 &10.53 &12.63 &14.41 &\multicolumn{1}{c}{11.46}\\
			
			\multicolumn{1}{c}{GaGNet(Pro.)}  &\Checkmark &5.94 &1.63 &0.19 &0.36 &\textbf{2.52} &\textbf{2.82} &\textbf{3.08} &\textbf{3.30} &\multicolumn{1}{c}{\textbf{2.93}}  &\textbf{68.60} &\textbf{76.53} &\textbf{82.13} &\textbf{86.20} &\multicolumn{1}{c}{\textbf{78.37}}  &\textbf{8.91} &\textbf{11.15} &\textbf{13.23} &\textbf{15.03} &\multicolumn{1}{c}{\textbf{12.08}}\\
			\hline
	\end{tabular}}
	\label{tbl:unseen-babble-factory1-objective}
	\vspace{-0.3cm}
\end{table*}

\begin{figure}[t]
	\centering
	\centerline{\includegraphics[width=0.70\columnwidth]{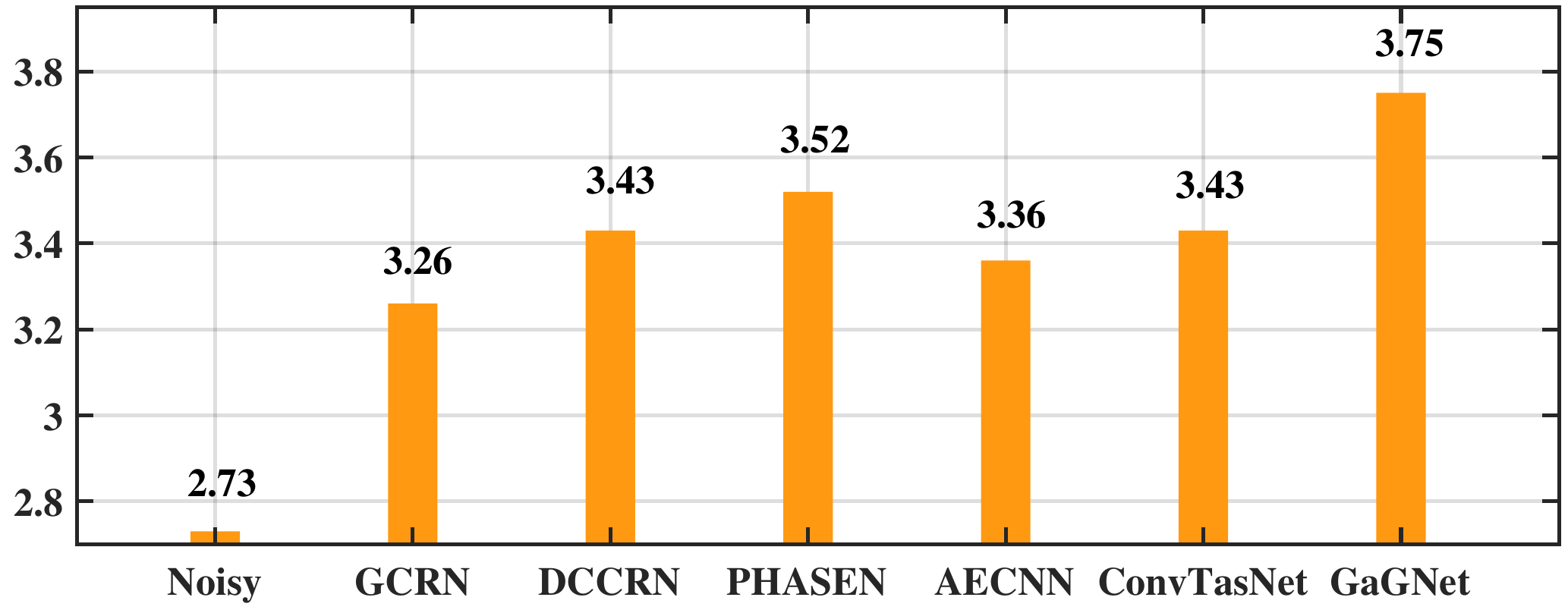}}
	\caption{DNSMOS comparison among different systems.}
	\label{fig:dnsmos}
	\vspace{-0.4cm}
\end{figure}

\renewcommand\arraystretch{0.7}
\begin{table}[t]
	\caption{Comparison with other state-of-the-art systems on the DNS Challenge non-blind test set.}
	\footnotesize
	\setlength{\tabcolsep}{2pt}
	\centering
%\resizebox{\columnwidth}{!}{
	\begin{tabular}{cccccc}
		\hline
		Models &Year &WB-PESQ &PESQ &STOI(\%) &SI-SDR(dB)\\
		\hline
		Noisy  &- &1.58 &2.45 &91.52 &9.07\\
		NSNet~{\cite{reddy2020interspeech}}  &2020 &2.15 &2.87  &94.47 &15.61\\
		DTLN~{\cite{westhausen2020dual}}  &2020 &- &3.04 &94.76  &16.34\\
		DCCRN~{\cite{hu2020dccrn}}  &2020 &- &3.27 &- &- \\
		PoCoNet~{\cite{isik2020poconet}} &2020 &2.75 &- &- &-\\
		FullSubNet~{\cite{hao2020fullsubnet}} &2021 &2.78  &3.31 &96.11 &17.29\\
		TRU-Net~{\cite{choi2021real}} &2021 &2.86  &3.36 &96.32 &17.55\\
		DCCRN+~{\cite{lv2021dccrn+}} &2021 &- &3.33 &- &- \\
		CTS-Net~{\cite{li2021icassp}} &2021 &2.94 &3.42  &96.66  &17.99\\
		GaGNet(Pro.)  &2021 &\textbf{3.17} &\textbf{3.56} &\textbf{97.13} &\textbf{18.91}\\
		\hline
	\end{tabular}
\label{tbl:dns1}
\vspace{-0.5cm}
\end{table}

\vspace{-0.2cm}
\section{Results and Analysis}
\label{results-and-analysis}
\subsection{Ablation Study}
\vspace{-0.1cm}
We randomly sample around 50,000 pairs from the training set to conduct the ablation study.
\subsubsection{The effect of $\left(P, Q\right)$}
We first investigate the effect \emph{w.r.t.} the number of TCM groups $P$ and GGMs $Q$. Three objective metrics are employed, namely perceptual evaluation of speech quality (PESQ)~{\cite{rix2001perceptual}}, extended short-time objective intelligibility (ESTOI)~{\cite{jensen2016algorithm}}, and signal-distortion ratio (SDR)~{\cite{vincent2007first}}. The number of parameters are also presented. The results are shown in Table~{\ref{tbl:ablation1}}. One can find that the increase of \textit{P} leads to improved scores, indicating that more TCMs can facilitate more accurate spectral estimation. Besides, when we increase the number of stages \textit{Q}, we also observe notable improvements in three metrics. It illustrates the advantage of multi-stage training, where the optimization target can be iteratively updated in a stage-wise manner. We set $\left(\textit{P}, \textit{Q} \right) = \left(2, 3\right)$ as the default configuration below, which can well balance between network performance and parameter burden.
%\vspace{-0.2cm}
\subsubsection{The effect of CRM}
We compare three different reconstruction algorithms to analyze the efficacy of the proposed CRM. In the first set, we drop the gaze block and only magnitude is considered, which is dubbed \textit{Mag-RM}. The second set neglects the glance block and enforces the gaze path to recover the overall complex spectrum from scratch, which is dubbed \textit{Com-RM}. Similar to PHASEN~{\cite{yin2020phasen}}, in the third set, the gaze block only contains the phase information and estimates the sine and cosine representations of the phase, which is dubbed \textit{Phasen-RM}. The results are presented in Table~{\ref{tbl:ablation2}}. First, \textit{Mag-RM} yields the worst score because of the phase information not being considered, which heavily hampers the speech quality. Second, \textit{Phasen-RM} is overall inferior to \textit{Com-RM}. This is because in the \textit{Phasen-RM}, phase is considered independently while \textit{Com-RM} jointly considers magnitude and phase by coupling them into the RI format. As phase usually exhibits rather random distribution, it is quite difficult for DNNs to predict the phase accurately. Moreover, the authors in~{\cite{yin2020phasen}} acknowledged that phase prediction is quite sensitive toward the nonlinear transform. Third, although relatively more parameters are needed for the proposed CRM , it achieves the best performance, which shows the superiority of spectral reconstruction with a collaborative method.

%\begin{figure}[t]
%	\setlength{\abovecaptionskip}{0.235cm}
%	\setlength{\belowcaptionskip}{-0.1cm}
%	\centering
%	\subfigure[Noisy magnitude spectrum]{
%		\includegraphics[width=0.46\columnwidth]{figures/noisy_mag_jet.pdf}	
%	}
%	\subfigure[Filtered magnitude by glance block]{
%		\includegraphics[width=0.46\columnwidth]{figures/glance_mag_jet.pdf}	
%	}
%	\subfigure[Estimated real detail by focus block]{
%		\includegraphics[width=0.46\columnwidth]{figures/focus_real_resi_jet.pdf}	
%	}
%	\subfigure[Estimated imaginary detail by focus block]{
%		\includegraphics[width=0.46\columnwidth]{figures/focus_imag_resi_jet.pdf}	
%	}
%	\caption{Visualization of the estimated spectra by glance and focus block. (a) Input noisy speech magnitude. (b) Filtered magnitude by glance block. (c) Estimated real spectral detail by focus block. (d) Estimated imaginary spectral detail by focus block.}
%	\label{fig:result-stage}
%	\vspace{-0.4cm}
%\end{figure}
\vspace{-0.4cm}
\subsection{Result Comparison with Baselines}
\vspace{-0.1cm}
Now we compare the objective performance of the proposed GaGNet with five advanced baselines, whose results are presented in Table~{\ref{tbl:unseen-babble-factory1-objective}}. Note that superscript ``$\dagger$'' denotes the case when the REL is replaced by the plain 2D-GLU to validate the effectiveness of REL module. Several observations can be made. First, when REL is adopted, notable improvements can be achieved than that with 2D-GLU. For example, going from GaGNet$^{\dagger}$ to GaGNet, average 0.11, 2.12\%, and 0.62dB improvements are observed in terms of PESQ, ESTOI, and SDR, respectively. This is because the introduction of UNet-Block can further exploit different feature scales and meanwhile recalibrate the feature distribution by residual connection. Second, the proposed system consistently surpasses other advanced baselines for different cases. For example, from DCCRN to GaGNet, average 0.21, 7.37\%, and 1.90\rm{dB} improvements are achieved with respect to PESQ, ESTOI, and SDR, respectively. This is because different from the previous direct complex spectral mapping strategy, a collaborative algorithm is proposed to reconstruct the target from complementary aspects, \emph{i.e.}, the magnitude information is coarsely estimated by the glance path, and meanwhile, the detail is further repaired by the gaze step. As both two paths are applied simultaneously, the pre-estimation error can be effectively mitigated. Besides, we follow the multi-stage learning pipeline and iteratively facilitate the spectrum refinement. It is noteworthy that our model still surpasses two non-causal baselines, \emph{i.e.} PHASEN and AECNN, despite only the past information can be exploited, which further reveals the superiority of our system. 

We also provide the number of parameters, multiply-accumulate operations (MACs), processing time, and memory footprint in the backward pass of different models, as shown in Table~{\ref{tbl:unseen-babble-factory1-objective}}. Note that both MACs and  processing time are calculated given the input audio of 16000 samples (1 second) on an Intel Core(TM) i5-4300 CPU clocked at 1.90GHz. The memory footprint is evaluated with a batch size of 4 on NVIDIA GeForce GTX 1660Ti. One can observe that the proposed system has overall lower statistics compared with other T-F domain based peers. This is because we only utilize the 2D convolutions in the encoder and abandon the widely-used 2D deconvolution-based decoder structure whose computational complexity is usually unbearable. Note that despite ConvTasNet utilizes an extremely small frame length (around 2ms), leading to low frame-level processing latency, as more frames are needed to process, the overall computation load is still large.

To demonstrate the merit of the proposed system in subjective quality, the DNSMOS~{\cite{reddy2020dnsmos}} is adopted as the criterion, which estimates subjective human ratings well. In total, 750 noisy utterances are generated and then processed by the SE systems. The results are shown in Fig.~{\ref{fig:dnsmos}}. One can see that the proposed approach yields the highest score, which verifies the superiority of the proposed system in subjective perception of speech quality.
\vspace{-0.3cm}
\renewcommand\arraystretch{0.7}
\begin{table}[t]
	\caption{Comparison with other state-of-the-art systems on the VoiceBank+Demand dataset.}
	\footnotesize
	\setlength{\tabcolsep}{2pt}
	\centering
	%\resizebox{\columnwidth}{!}{
	\begin{tabular}{cccccc}
		\hline
		Models &Year &WB-PESQ &CSIG &CBAK &COVL\\
		\hline
		Noisy  &- &1.97 &3.35 &2.44 &2.63\\
		SEGAN~{\cite{pascual2017segan}}  &2017 &2.16 &3.48  &2.94 &2.80\\
		MetricGAN~{\cite{metricgan}}  &2019 &2.86 &3.99 &3.18 &3.42\\
		MHSA+SPk~{\cite{MHSA+SPk}}  &2020 &2.99 &4.15 &3.42 &3.57 \\
		PHASEN~{\cite{yin2020phasen}} &2020 &2.99 &4.21 &\textbf{3.55} &3.62\\
		DEMUCS~{\cite{defossez2020real}} &2021 &3.07  &\textbf{4.31} &3.40 &3.63\\
		DCCRN+~{\cite{lv2021dccrn+}} &2021 &2.84 &- &- &- \\
		MetricGAN+~{\cite{metricgan+}} &2021 &\textbf{3.15} &4.14 &3.16 &\textbf{3.64} \\
		GaGNet(Pro.)  &2021 &2.94 &4.26 &3.45 &3.59\\
		\hline
	\end{tabular}
	\label{tbl:vb}
	\vspace{-0.5cm}
\end{table}
\vspace{-0.1cm}
\subsection{Comparison on other benchmarks}
\vspace{-0.0cm}
Except for WSJ0-SI84, we also conduct experiments on two benchmarks, namely DNS-Challenge dataset and Voicebank+Demand dataset~{\cite{pascual2017segan}}. For DNS-Challenge dataset, we totally generate around 3000 hours noisy-clean pairs for training with the provided clean utterances and noise sets. The non-blind test set with 150 pairs is adopted for model evaluation. Four metrics are utilized, namely wide-band PESQ (WB-PESQ)~{\cite{rec2005p}}, PESQ, STOI~{\cite{jensen2016algorithm}}, and SI-SDR. For Voicebank+Demand dataset, 11,572 and 824 pairs are for training and testing. Four metrics are utilized, namely WB-PESQ, CSIG, CBAK, and COVL~{\cite{cmos}}. Table~{\ref{tbl:dns1}} shows the comparison with state-of-the-art (SOTA) approaches on the DNS-Challenge dataset. One can find that our system outperforms previous baselines by a large margin for all four metrics. The results on Voicebank+Demand dataset are presented in Table~{\ref{tbl:vb}}, one can get that despite not obtaining the highest score, our system still achieves competitive results compared with previous SOTA systems. We attribute the reason as the SNR level of the testing samples in the Voicebank+Demand dataset is rather high. Therefore, repeatedly stacking GGM can not bring a notable performance advantage. Besides, only the MSE is adopted as the loss, leading to a limited improvement on the specific metric like PESQ. We provide the processed samples, which are available at https://github.com/Andong-Li-speech/GaGNet.
\vspace{-0.3cm}
\section{Conclusion}
\vspace{-0.1cm}
We present a collaborative framework to tackle monaural speech enhancement in the complex domain. Specifically, we decouple the complex spectrum optimization problem and propose a dual-path reconstruction algorithm, where the glance path seeks to estimate the overall spectrum in the magnitude domain, and the gaze path is tasked with repairing the missing spectral detail in the complex domain. The two paths are applied simultaneously to facilitate the target recovery from different scales. Besides, the multi-stage training strategy is adopted, which enforces the target optimization progressively. We conduct extensive experiments on the WSJ0-SI84, DNS-Challenge dataset, and Voicebank+Demand dataset. The results show that our method yields promising performance over three datasets.
%\section*{Acknowledgment}
\renewcommand{\baselinestretch}{0.96}
\clearpage
%\section*{References}
\bibliographystyle{ieeetr}
%\bibliography{bibfile}

\begin{thebibliography}{10}

\bibitem{loizou2013speech}
P.~C. Loizou, {\em Speech enhancement: theory and practice}.
\newblock CRC press, 2013.

\bibitem{wang2018supervised}
D.~Wang and J.~Chen, ``Supervised speech separation based on deep learning: An
  overview,'' {\em IEEE/ACM Transactions on Audio, Speech, and Language
  Processing}, vol.~26, no.~10, pp.~1702--1726, 2018.

\bibitem{xu2013experimental}
Y.~Xu, J.~Du, L.-R. Dai, and C.-H. Lee, ``An experimental study on speech
  enhancement based on deep neural networks,'' {\em IEEE Signal processing
  letters}, vol.~21, no.~1, pp.~65--68, 2013.

\bibitem{kolbaek2016speech}
M.~Kolb{\ae}k, Z.-H. Tan, and J.~Jensen, ``Speech intelligibility potential of
  general and specialized deep neural network based speech enhancement
  systems,'' {\em IEEE/ACM Transactions on Audio, Speech, and Language
  Processing}, vol.~25, no.~1, pp.~153--167, 2016.

\bibitem{ephraim1984speech}
Y.~Ephraim and D.~Malah, ``Speech enhancement using a minimum-mean square error
  short-time spectral amplitude estimator,'' {\em IEEE Transactions on
  acoustics, speech, and signal processing}, vol.~32, no.~6, pp.~1109--1121,
  1984.

\bibitem{tan2018gated}
K.~Tan, J.~Chen, and D.~Wang, ``Gated residual networks with dilated
  convolutions for monaural speech enhancement,'' {\em IEEE/ACM transactions on
  audio, speech, and language processing}, vol.~27, no.~1, pp.~189--198, 2018.

\bibitem{xu2020using}
Z.~Xu, S.~Elshamy, and T.~Fingscheidt, ``Using separate losses for speech and
  noise in mask-based speech enhancement,'' in {\em ICASSP 2020-2020 IEEE
  International Conference on Acoustics, Speech and Signal Processing
  (ICASSP)}, pp.~7519--7523, IEEE, 2020.

\bibitem{gerkmann2015phase}
T.~Gerkmann, M.~Krawczyk-Becker, and J.~Le~Roux, ``Phase processing for
  single-channel speech enhancement: History and recent advances,'' {\em IEEE
  signal processing Magazine}, vol.~32, no.~2, pp.~55--66, 2015.

\bibitem{mowlaee2016advances}
P.~Mowlaee, R.~Saeidi, and Y.~Stylianou, ``Advances in phase-aware signal
  processing in speech communication,'' {\em Speech communication}, vol.~81,
  pp.~1--29, 2016.

\bibitem{paliwal2011importance}
K.~Paliwal, K.~W{\'o}jcicki, and B.~Shannon, ``The importance of phase in
  speech enhancement,'' {\em speech communication}, vol.~53, no.~4,
  pp.~465--494, 2011.

\bibitem{williamson2015complex}
D.~S. Williamson, Y.~Wang, and D.~Wang, ``Complex ratio masking for monaural
  speech separation,'' {\em IEEE/ACM transactions on audio, speech, and
  language processing}, vol.~24, no.~3, pp.~483--492, 2015.

\bibitem{tan2020learning}
K.~Tan and D.~Wang, ``Learning complex spectral mapping with gated
  convolutional recurrent networks for monaural speech enhancement,'' {\em
  IEEE/ACM Transactions on Audio, Speech, and Language Processing}, vol.~28,
  pp.~380--390, 2020.

\bibitem{lee2019joint}
J.~Lee and H.-G. Kang, ``A joint learning algorithm for complex-valued tf masks
  in deep learning-based single-channel speech enhancement systems,'' {\em
  IEEE/ACM Transactions on Audio, Speech, and Language Processing}, vol.~27,
  no.~6, pp.~1098--1108, 2019.

\bibitem{pascual2017segan}
S.~Pascual, A.~Bonafonte, and J.~Serr{\`a}, ``Segan: Speech enhancement
  generative adversarial network,'' {\em Proc. Interspeech 2017},
  pp.~3642--3646, 2017.

\bibitem{luo2019conv}
Y.~Luo and N.~Mesgarani, ``Conv-tasnet: Surpassing ideal time--frequency
  magnitude masking for speech separation,'' {\em IEEE/ACM transactions on
  audio, speech, and language processing}, vol.~27, no.~8, pp.~1256--1266,
  2019.

\bibitem{pandey2019new}
A.~Pandey and D.~Wang, ``A new framework for cnn-based speech enhancement in
  the time domain,'' {\em IEEE/ACM Transactions on Audio, Speech, and Language
  Processing}, vol.~27, no.~7, pp.~1179--1188, 2019.

\bibitem{le2019sdr}
J.~Le~Roux, S.~Wisdom, H.~Erdogan, and J.~R. Hershey, ``Sdr--half-baked or well
  done?,'' in {\em ICASSP 2019-2019 IEEE International Conference on Acoustics,
  Speech and Signal Processing (ICASSP)}, pp.~626--630, IEEE, 2019.

\bibitem{reddy2020dnsmos}
C.~K. Reddy, V.~Gopal, and R.~Cutler, ``Dnsmos: A non-intrusive perceptual
  objective speech quality metric to evaluate noise suppressors,'' {\em arXiv
  preprint arXiv:2010.15258}, 2020.

\bibitem{fayek2020progressive}
H.~M. Fayek, L.~Cavedon, and H.~R. Wu, ``Progressive learning: A deep learning
  framework for continual learning,'' {\em Neural Networks}, vol.~128,
  pp.~345--357, 2020.

\bibitem{li2020speech}
A.~Li, M.~Yuan, C.~Zheng, and X.~Li, ``Speech enhancement using progressive
  learning-based convolutional recurrent neural network,'' {\em Applied
  Acoustics}, vol.~166, p.~107347, 2020.

\bibitem{li2021icassp}
A.~Li, W.~Liu, X.~Luo, C.~Zheng, and X.~Li, ``Icassp 2021 deep noise
  suppression challenge: Decoupling magnitude and phase optimization with a
  two-stage deep network,'' {\em arXiv preprint arXiv:2102.04198}, 2021.

\bibitem{hao2020masking}
X.~Hao, X.~Su, S.~Wen, Z.~Wang, Y.~Pan, F.~Bao, and W.~Chen, ``Masking and
  inpainting: A two-stage speech enhancement approach for low snr and
  non-stationary noise,'' in {\em ICASSP 2020-2020 IEEE International
  Conference on Acoustics, Speech and Signal Processing (ICASSP)},
  pp.~6959--6963, IEEE, 2020.

\bibitem{wang2020glance}
Y.~Wang, K.~Lv, R.~Huang, S.~Song, L.~Yang, and G.~Huang, ``Glance and focus: a
  dynamic approach to reducing spatial redundancy in image classification,''
  {\em Advances in Neural Information Processing Systems}, vol.~33, 2020.

\bibitem{braun2021towards}
S.~Braun, H.~Gamper, C.~K. Reddy, and I.~Tashev, ``Towards efficient models for
  real-time deep noise suppression,'' {\em arXiv preprint arXiv:2101.09249},
  2021.

\bibitem{qin2020u2}
X.~Qin, Z.~Zhang, C.~Huang, M.~Dehghan, O.~R. Zaiane, and M.~Jagersand,
  ``U2-net: Going deeper with nested u-structure for salient object
  detection,'' {\em Pattern Recognition}, vol.~106, p.~107404, 2020.

\bibitem{ulyanov2016instance}
D.~Ulyanov, A.~Vedaldi, and V.~Lempitsky, ``Instance normalization: The missing
  ingredient for fast stylization,'' {\em arXiv preprint arXiv:1607.08022},
  2016.

\bibitem{he2015delving}
K.~He, X.~Zhang, S.~Ren, and J.~Sun, ``Delving deep into rectifiers: Surpassing
  human-level performance on imagenet classification,'' in {\em Proc. of ICCV},
  pp.~1026--1034, 2015.

\bibitem{dauphin2017language}
Y.~N. Dauphin, A.~Fan, M.~Auli, and D.~Grangier, ``Language modeling with gated
  convolutional networks,'' in {\em International conference on machine
  learning}, pp.~933--941, PMLR, 2017.

\bibitem{zhang2020deepmmse}
Q.~Zhang, A.~Nicolson, M.~Wang, K.~K. Paliwal, and C.~Wang, ``Deepmmse: A deep
  learning approach to mmse-based noise power spectral density estimation,''
  {\em IEEE/ACM Transactions on Audio, Speech, and Language Processing},
  vol.~28, pp.~1404--1415, 2020.

\bibitem{wang2020complex}
Z.-Q. Wang, P.~Wang, and D.~Wang, ``Complex spectral mapping for single-and
  multi-channel speech enhancement and robust asr,'' {\em IEEE/ACM Transactions
  on Audio, Speech, and Language Processing}, vol.~28, pp.~1778--1787, 2020.

\bibitem{wisdom2019differentiable}
S.~Wisdom, J.~R. Hershey, K.~Wilson, J.~Thorpe, M.~Chinen, B.~Patton, and R.~A.
  Saurous, ``Differentiable consistency constraints for improved deep speech
  enhancement,'' in {\em ICASSP 2019-2019 IEEE International Conference on
  Acoustics, Speech and Signal Processing (ICASSP)}, pp.~900--904, IEEE, 2019.

\bibitem{paul1992design}
D.~Paul and J.~Baker, ``The design for the wall street journal-based csr
  corpus,'' in {\em Workshop on Speech and Natural Language}, p.~357–362,
  1992.

\bibitem{reddy2020interspeech}
C.~K. Reddy, V.~Gopal, R.~Cutler, E.~Beyrami, R.~Cheng, H.~Dubey,
  S.~Matusevych, R.~Aichner, A.~Aazami, S.~Braun, {\em et~al.}, ``The
  interspeech 2020 deep noise suppression challenge: Datasets, subjective
  testing framework, and challenge results,'' {\em Proc. Interspeech 2020},
  pp.~2492--2496, 2020.

\bibitem{varga1993assessment}
A.~Varga and H.~Steeneken, ``Assessment for automatic speech recognition: Ii.
  noisex-92: A database and an experiment to study the effect of additive noise
  on speech recognition systems,'' {\em Speech Commun.}, vol.~12, no.~3,
  pp.~247--251, 1993.

\bibitem{hu2020dccrn}
Y.~Hu, Y.~Liu, S.~Lv, M.~Xing, S.~Zhang, Y.~Fu, J.~Wu, B.~Zhang, and L.~Xie,
  ``Dccrn: Deep complex convolution recurrent network for phase-aware speech
  enhancement,'' {\em Proc. Interspeech 2020}, pp.~2472--2476, 2020.

\bibitem{yin2020phasen}
D.~Yin, C.~Luo, Z.~Xiong, and W.~Zeng, ``Phasen: A phase-and-harmonics-aware
  speech enhancement network,'' in {\em Proceedings of the AAAI Conference on
  Artificial Intelligence}, vol.~34, pp.~9458--9465, 2020.

\bibitem{li2021importance}
A.~Li, C.~Zheng, R.~Peng, and X.~Li, ``On the importance of power compression
  and phase estimation in monaural speech dereverberation,'' {\em JASA Express
  Letters}, vol.~1, no.~1, p.~014802, 2021.

\bibitem{kingma2014adam}
D.~Kingma and J.~Ba, ``Adam: A method for stochastic optimization,'' {\em arXiv
  preprint arXiv:1412.6980}, 2014.

\bibitem{westhausen2020dual}
N.~L. Westhausen and B.~T. Meyer, ``Dual-signal transformation lstm network for
  real-time noise suppression,'' {\em Proc. Interspeech 2020}, pp.~2477--2481,
  2020.

\bibitem{hao2020fullsubnet}
X.~Hao, X.~Su, R.~Horaud, and X.~Li, ``Fullsubnet: A full-band and sub-band
  fusion model for real-time single-channel speech enhancement,'' {\em arXiv
  preprint arXiv:2010.15508}, 2020.

\bibitem{choi2021real}
H.-S. Choi, S.~Park, J.~H. Lee, H.~Heo, D.~Jeon, and K.~Lee, ``Real-time
  denoising and dereverberation with tiny recurrent u-net,'' {\em arXiv
  preprint arXiv:2102.03207}, 2021.

\bibitem{isik2020poconet}
U.~Isik, R.~Giri, N.~Phansalkar, J.-M. Valin, K.~Helwani, and A.~Krishnaswamy,
  ``Poconet: Better speech enhancement with frequency-positional embeddings,
  semi-supervised conversational data, and biased loss,'' {\em Proc.
  Interspeech 2020}, pp.~2487--2491, 2020.

\bibitem{rix2001perceptual}
A.~Rix, J.~Beerends, M.~Hollier, and A.~Hekstra, ``Perceptual evaluation of
  speech quality ({PESQ})-a new method for speech quality assessment of
  telephone networks and codecs,'' in {\em Proc. of ICASSP}, vol.~2,
  pp.~749--752, IEEE, 2001.

\bibitem{jensen2016algorithm}
J.~Jensen and C.~Taal, ``An algorithm for predicting the intelligibility of
  speech masked by modulated noise maskers,'' {\em IEEE/ACM Trans.~Audio Speech
  Lang.~Proc.}, vol.~24, no.~11, pp.~2009--2022, 2016.

\bibitem{vincent2007first}
E.~Vincent, H.~Sawada, P.~Bofill, S.~Makino, and J.~Rosca, ``First stereo audio
  source separation evaluation campaign: data, algorithms and results,'' in
  {\em International Conference on Independent Component Analysis and Signal
  Separation}, pp.~552--559, Springer, 2007.

\bibitem{rec2005p}
I.~Rec, ``P. 862.2: Wideband extension to recommendation p. 862 for the
  assessment of wideband telephone networks and speech codecs,'' {\em
  International Telecommunication Union, CH--Geneva}, 2005.

\end{thebibliography}


\begin{thebibliography}{10}
	\bibitem{loizou2013speech}
	P.~C. Loizou, {\em Speech enhancement: theory and practice}.
	\newblock CRC press, 2013.
	
	\bibitem{wang2018supervised}
	D.~Wang and J.~Chen, ``Supervised speech separation based on deep learning: An
	overview,'' {\em IEEE/ACM Trans.~Audio.~Speech, Lang.~Process.}, vol.~26, no.~10, pp.~1702--1726, 2018.
	
	\bibitem{xu2013experimental}
	Y.~Xu, J.~Du, L.-R. Dai, and C.-H. Lee, ``An experimental study on speech
	enhancement based on deep neural networks,'' {\em IEEE Signal Process. Lett.}, vol.~21, no.~1, pp.~65--68, 2013.
	
	\bibitem{siniscalchi2021vector}
	S.M.~Siniscalchi, ``Vector-to-Vector Regression via Distributional Loss for Speech Enhancement,'' {\em IEEE Signal Process. Lett.}, vol.~28, pp.~254--258, 2021.
		
	\bibitem{ephraim1984speech}
	Y.~Ephraim and D.~Malah, ``Speech enhancement using a minimum-mean square error short-time spectral amplitude estimator,'' {\em IEEE Trans.~Acoust., Speech, Signal Process.}, vol.~32, no.~6, pp.~1109--1121, 1984.
	
	\bibitem{tan2018gated}
	K.~Tan, J.~Chen, and D.~Wang, ``Gated residual networks with dilated
	convolutions for monaural speech enhancement,'' {\em IEEE/ACM Trans.~Audio.~Speech, Lang.~Process.}, vol.~27, no.~1, pp.~189--198, 2018.
	
	\bibitem{gerkmann2015phase}
	T.~Gerkmann, M.~Krawczyk-Becker, and J.~Le~Roux, ``Phase processing for
	single-channel speech enhancement: History and recent advances,'' {\em IEEE Signal Process. Mag.}, vol.~32, no.~2, pp.~55--66, 2015.
	
	\bibitem{mowlaee2016advances}
	P.~Mowlaee, R.~Saeidi, and Y.~Stylianou, ``Advances in phase-aware signal processing in speech communication,'' {\em Speech Commun.}, vol.~81,
	pp.~1--29, 2016.
	
	\bibitem{paliwal2011importance}
	K.~Paliwal, K.~W{\'o}jcicki, and B.~Shannon, ``The importance of phase in
	speech enhancement,'' {\em Speech Commun.}, vol.~53, no.~4,
	pp.~465--494, 2011.
	
	\bibitem{williamson2015complex}
	D.~S. Williamson, Y.~Wang, and D.~Wang, ``Complex ratio masking for monaural
	speech separation,'' {\em IEEE/ACM Trans.~Audio.~Speech, Lang.~Process.}, vol.~24, no.~3, pp.~483--492, 2015.
	
	\bibitem{tan2020learning}
	K.~Tan and D.~Wang, ``Learning complex spectral mapping with gated
	convolutional recurrent networks for monaural speech enhancement,'' {\em
		IEEE/ACM Trans.~Audio.~Speech, Lang.~Process.}, vol.~28,
	pp.~380--390, 2020.
	
	\bibitem{lee2019joint}
	J.~Lee and H.-G. Kang, ``A joint learning algorithm for complex-valued tf masks
	in deep learning-based single-channel speech enhancement systems,'' {\em
		IEEE/ACM Trans.~Audio.~Speech, Lang.~Process.}, vol.~27,
	no.~6, pp.~1098--1108, 2019.
	
	\bibitem{pascual2017segan}
	S.~Pascual, A.~Bonafonte, and J.~Serr{\`a}, ``SEGAN: Speech Enhancement Generative Adversarial Network,'' in {\em Proc. Interspeech 2017},
	pp.~3642--3646, 2017.
	
	\bibitem{luo2019conv}
	Y.~Luo and N.~Mesgarani, ``Conv-tasnet: Surpassing ideal time--frequency magnitude masking for speech separation,'' {\em IEEE/ACM Trans.~Audio.~Speech, Lang.~Process.}, vol.~27, no.~8, pp.~1256--1266,
	2019.
	
	\bibitem{pandey2019new}
	A.~Pandey and D.~Wang, ``A new framework for CNN-based speech enhancement in the time domain,'' {\em IEEE/ACM Trans.~Audio.~Speech, Lang.~Process.}, vol.~27, no.~7, pp.~1179--1188, 2019.
	
	\bibitem{defossez2020real}
	A.~Defossez, G.~Synnaeve, and Y.~Adi, ``Real Time Speech Enhancement in the Waveform Domain,'' in {\em Proc. Interspeech 2020}, pp.~3291--3295, 2020.
	
	\bibitem{le2019sdr}
	J.~Le~Roux, S.~Wisdom, H.~Erdogan, and J.~R. Hershey, ``SDR--half-baked or well
	done?,'' in {\em Proc.~ICASSP 2019}, pp.~626--630, IEEE, 2019.
	
	\bibitem{reddy2020dnsmos}
	C.~K. Reddy, V.~Gopal, and R.~Cutler, ``DNSMOS: A Non-Intrusive Perceptual Objective Speech Quality metric to evaluate Noise Suppressors,'' {\em arXiv preprint arXiv:2010.15258}, 2020.
	
	\bibitem{gao2016snr}
	T.~Gao, J.~Du, L.-R.~Dai, and C.-H.~Lee, ``SNR-Based Progressive Learning of Deep Neural Network for Speech Enhancement,'' in {\em Proc.~Interspeech 2016}, pp.~3713--3717, 2016.

	
	\bibitem{li2021icassp}
	A.~Li, W.~Liu, C.~Zheng, and X.~Li, ``Two Heads are Better Than One: A Two-Stage Complex Spectral Mapping Approach for Monaural Speech Enhancement,'' {\em IEEE/ACM Trans.~Audio.~Speech, Lang.~Process.}, vol.~29, pp.~1829--1843, 2021.
	
	\bibitem{hao2020masking}
	X.~Hao, X.~Su, S.~Wen, Z.~Wang, Y.~Pan, F.~Bao, and W.~Chen, ``Masking and
	inpainting: A two-stage speech enhancement approach for low snr and
	non-stationary noise,'' in {\em Proc.~ICASSP 2020},
	pp.~6959--6963, IEEE, 2020.
	
	\bibitem{wang2020glance}
	Y.~Wang, K.~Lv, R.~Huang, S.~Song, L.~Yang, and G.~Huang, ``Glance and focus: a
	dynamic approach to reducing spatial redundancy in image classification,''in
	{\em Proc.~NeurIPS}, vol.~33, 2020.
	
	\bibitem{qin2020u2}
	X.~Qin, Z.~Zhang, C.~Huang, M.~Dehghan, O.~R. Zaiane, and M.~Jagersand,
	``U2-Net: Going deeper with nested U-structure for salient object detection,'' {\em Pattern Recognit.}, vol.~106, p.~107404, 2020.
	
	\bibitem{dauphin2017language}
	Y.~N. Dauphin, A.~Fan, M.~Auli, and D.~Grangier, ``Language modeling with gated
	convolutional networks,'' in {\em Proc.~ICML}, pp.~933--941, PMLR, 2017.
	
	\bibitem{he2015delving}
	K.~He, X.~Zhang, S.~Ren, and J.~Sun, ``Delving deep into rectifiers: Surpassing human-level performance on imagenet classification,'' in {\em Proc. of ICCV},
	pp.~1026--1034, 2015.
	
	\bibitem{zhang2020deepmmse}
	Q.~Zhang, A.~Nicolson, M.~Wang, K.~K. Paliwal, and C.~Wang, ``Deepmmse: A deep
	learning approach to MMSE-based noise power spectral density estimation,''
	{\em IEEE/ACM Trans.~Audio.~Speech, Lang.~Process.},
	vol.~28, pp.~1404--1415, 2020.
	
	\bibitem{wang2020complex}
	Z.-Q. Wang, P.~Wang, and D.~Wang, ``Complex spectral mapping for single-and multi-channel speech enhancement and robust ASR,'' {\em IEEE Trans.~Acoust., Speech, Signal Process.}, vol.~28, pp.~1778--1787, 2020.
	
	\bibitem{wisdom2019differentiable}
	S.~Wisdom, J.~R. Hershey, K.~Wilson, J.~Thorpe, M.~Chinen, B.~Patton, and R.~A.
	Saurous, ``Differentiable consistency constraints for improved deep speech
	enhancement,'' in {\em Proc.~ICASSP 2019}, pp.~900--904, IEEE, 2019.
	
	\bibitem{paul1992design}
	D.~Paul and J.~Baker, ``The design for the Wall Street Journal-based CSR corpus,'' in {\em Proc.~Workshop on Speech and Natural Lang.}, p.~357–362, 1992.
	
	\bibitem{reddy2020interspeech}
	C.~K. Reddy, V.~Gopal, R.~Cutler, E.~Beyrami, R.~Cheng, H.~Dubey,
	S.~Matusevych, R.~Aichner, A.~Aazami, S.~Braun, {\em et~al.}, ``The INTERSPEECH 2020 Deep Noise Suppression Challenge: Datasets, Subjective Testing Framework, and Challenge Results,'' in {\em Proc. Interspeech 2020},
	pp.~2492--2496, 2020.
	
	\bibitem{varga1993assessment}
	A.~Varga and H.~Steeneken, ``Assessment for automatic speech recognition: II. NOISEX-92: A database and an experiment to study the effect of additive noise on speech recognition systems,'' {\em Speech Commun.}, vol.~12, no.~3,
	pp.~247--251, 1993.
	
	\bibitem{hu2020dccrn}
	Y.~Hu, Y.~Liu, S.~Lv, M.~Xing, S.~Zhang, Y.~Fu, J.~Wu, B.~Zhang, and L.~Xie,
	``DCCRN: Deep Complex Convolution Recurrent Network for Phase-Aware Speech Enhancement,'' in {\em Proc. Interspeech 2020}, pp.~2472--2476, 2020.
	
	\bibitem{lv2021dccrn+}
	S.~Lv, Y.~Hu, S.~Zhang, and L.~Xie, ``DCCRN+: Channel-wise Subband DCCRN with SNR Estimation for Speech Enhancement,'' {\em arXiv preprint arXiv:2106.08672}, 2021.
	
	\bibitem{yin2020phasen}
	D.~Yin, C.~Luo, Z.~Xiong, and W.~Zeng, ``Phasen: A phase-and-harmonics-aware
	speech enhancement network,'' in {\em Proc.~AAAI}, vol.~34, pp.~9458--9465, 2020.
	
	\bibitem{li2021importance}
	A.~Li, C.~Zheng, R.~Peng, and X.~Li, ``On the importance of power compression
	and phase estimation in monaural speech dereverberation,'' {\em JASA Express Letters}, vol.~1, no.~1, p.~014802, 2021.
	
	\bibitem{kingma2014adam}
	D.~Kingma and J.~Ba, ``Adam: A method for stochastic optimization,'' {\em arXiv preprint arXiv:1412.6980}, 2014.
	
	\bibitem{westhausen2020dual}
	N.~L. Westhausen and B.~T. Meyer, ``Dual-Signal Transformation LSTM Network for Real-Time Noise Suppression,'' in{\em Proc. Interspeech 2020}, pp.~2477--2481,
	2020.
	
	\bibitem{hao2020fullsubnet}
	X.~Hao, X.~Su, R.~Horaud, and X.~Li, ``FullSubNet: A Full-Band and Sub-Band Fusion Model for Real-Time Single-Channel Speech Enhancement,'' {\em arXiv preprint arXiv:2010.15508}, 2020.
	
	\bibitem{choi2021real}
	H.-S. Choi, S.~Park, J.~H. Lee, H.~Heo, D.~Jeon, and K.~Lee, ``Real-time
	denoising and dereverberation with tiny recurrent u-net,'' {\em arXiv
		preprint arXiv:2102.03207}, 2021.
	
	\bibitem{isik2020poconet}
	U.~Isik, R.~Giri, N.~Phansalkar, J.-M. Valin, K.~Helwani, and A.~Krishnaswamy,
	``PoCoNet: Better Speech Enhancement with Frequency-Positional Embeddings, Semi-Supervised Conversational Data, and Biased Loss,'' in {\em Proc.
		Interspeech 2020}, pp.~2487--2491, 2020.
	
	\bibitem{rix2001perceptual}
	A.~Rix, J.~Beerends, M.~Hollier, and A.~Hekstra, ``Perceptual evaluation of
	speech quality ({PESQ})-a new method for speech quality assessment of
	telephone networks and codecs,'' in {\em Proc.~ICASSP}, pp.~749--752, IEEE, 2001.
	
	\bibitem{jensen2016algorithm}
	J.~Jensen and C.~Taal, ``An algorithm for predicting the intelligibility of
	speech masked by modulated noise maskers,'' {\em IEEE/ACM Trans.~Audio Speech
		Lang.~Proc.}, vol.~24, no.~11, pp.~2009--2022, 2016.
	
	\bibitem{vincent2007first}
	E.~Vincent, H.~Sawada, P.~Bofill, S.~Makino, and J.~Rosca, ``First stereo audio
	source separation evaluation campaign: data, algorithms and results,'' in
	{\em International Conference on Independent Component Analysis and Signal
		Separation}, pp.~552--559, Springer, 2007.
	
	\bibitem{rec2005p}
	ITU-T Recommendation P.862.2,
	\newblock {\em Wideband extension to recommendation p. 862 for the assessment of wideband telephone networks and speech codecs}, 2007.
	
	\bibitem{MHSA+SPk}
	Y.~Koizumi, K.~Yaiabe, M.~Delcroix, Y.~Maxuxama, and D.~Takeuchi, ``Speech enhancement using self-adaptation and multihead self-attention," in {\em Proc.~ICASSP}, pp.~2472--2476, 2020.
	
	\bibitem{metricgan}
	S.-W.~Fu, C.-F.~Liao, Y.~Tsao, and S.-D.~Lin, ``Metricgan: Generative adversarial networks based black-box metric scores optimization for speech enhancement,'' in {\em Proc. ICML 2019}, pp.~2031--2041, 2019.
	
	\bibitem{metricgan+}
	S.-W.~Fu, C.~Yu, T.-A.~Hsieh, et.al., ``MetricGAN+: An Improved Version of MetricGAN for Speech Enhancement,'' {\em arXiv preprint arXiv:2104.03538}, 2021.
	
	\bibitem{cmos}
	Y.~Hu, and P.C.~Loizou, ``Evaluation of objective quality measures for
	speech enhancement,'' {\em IEEE/ACM Trans.~Audio.~Speech, Lang.~Process.}, vol.~16, no.~1, pp.~229--238, 2007.
	
\end{thebibliography}

\end{document}